\documentclass[%
 aip,
 amsmath,amssymb,
 reprint,%
]{revtex4-1}

\pdfoutput=1 

\usepackage{graphicx}
\usepackage{dcolumn}
\usepackage{bm}

\usepackage[utf8]{inputenc}
\usepackage[T1,T2A]{fontenc}
\usepackage{mathptmx}
\usepackage{etoolbox}
\usepackage{mhchem}

\usepackage{amsmath,amsfonts,amssymb,flafter,float,graphicx,latexsym,natbib,verbatim,xfrac}
\usepackage[russian, english]{babel}
\usepackage[usenames]{color}
\usepackage[mathscr]{euscript}
\usepackage[pdftex,colorlinks=true,linkcolor=blue,urlcolor=blue,citecolor=blue]{hyperref}

\makeatletter
\def\@email#1#2{%
 \endgroup
 \patchcmd{\titleblock@produce}
  {\frontmatter@RRAPformat}
  {\frontmatter@RRAPformat{\produce@RRAP{*#1\href{mailto:#2}{#2}}}\frontmatter@RRAPformat}
  {}{}
}%
\makeatother

\DeclareUnicodeCharacter{2212}{-}

\begin{document}

\preprint{AIP/123-QED}

\title[Tuning the conductance topology in solids]{Tuning the conductance topology in solids}
\author{Victor Lopez-Richard}
 \email{vlopez@df.ufscar.br}
 \altaffiliation{Physics Department, Federal University of São Carlos, 13565-905 São Carlos, SP, Brazil}
\author{Rafael Schio Wengenroth Silva}%
 \email{rafaelschio@df.ufscar.br}
\affiliation{ 
Physics Department, Federal University of São Carlos, 13565-905 São Carlos, SP, Brazil
}%

\author{Ovidiu Lipan}
 \email{olipan@richmond.edu}
\affiliation{Department of Physics, University of Richmond, 28 Westhampton Way, Richmond, Virginia 23173, USA
}%

\author{Fabian Hartmann}
 \email{fabian.hartmann@physik.uni-wuerzburg.de }
\affiliation{%
Technische Physik, Physikalisches Institut and Röntgen Center for Complex Material Systems (RCCM), Universität Würzburg, Am Hubland, D-97074 Würzburg, Germany
}%

\date{\today}

\begin{abstract}
The inertia of trapping and detrapping of nonequilibrium charge carriers affects the electrochemical and transport properties of both bulk and nanoscopic structures in a very peculiar way. An emerging memory response with a hysteresis in the current–voltage response and its eventual multiple crossing, produced by this universally available ingredient, are signatures of this process. Here, we deliver a microscopic and analytical solution for these behaviors, understood as the modulation of the topology of the current-voltage loops. The memory emergence becomes thus a characterization tool for intrinsic features that affect the electronic transport of solids such as the nature and number of trapping sites, intrinsic symmetry constraints, and natural relaxation time scales. This method is also able to reduce the seeming complexity of frequency-dependent electrochemical impedance and cyclic voltammetry observable for a variety of systems to a combination of simple microscopic ingredients.
\end{abstract}

\maketitle



\section{Introduction}

Resistive switching or memristive responses~\cite{Chua1971}, can be expected in various forms~\cite{Pershin2011}. They have recently become flexible building blocks for information storage, encryption tools~\cite{Lanza2022}, and advances have been foreseen for their applications in brain-inspired computing in the years to come~\cite{Kumar2022}. Even neuroprosthetic architectures are envisaged~\cite{Dias2022} due to their potential to mimic neuromorphic responses.
The resistive switching in non-linear systems is a relevant and pervasive effect that transcends memristive functionalities and may emerge in electrochemical processes, either intentionally or not. This is the case of ionic-controlled surface recombination processes detected in solar cells~\cite{Gonzales2022}, the memory traces of the impedance in pH-switchable polymer electrodes~\cite{MacVittie2013}, hydride formation along with hydrogen evolution reaction in semiconductor surfaces~\cite{Priyadarshani2021}, redox reactions that tune the conductance and capacitive properties of composite elements used for energy storage architectures~\cite{Luo2022}, redox-kinetics tuned by moisture affecting oxide surfaces~\cite{Messerschmitt2015}, among others. In all these cases, cyclic voltammetry and/or impedance spectroscopy~\cite{Shangshang2021} appear as complementary tools for the characterization of this peculiar transport response.

Various mechanisms are responsible for the memory emergence such as field-induced ion drifting, that leads to the formation of conductive paths of filamentary nature~\cite{Waser2012}, or volatile nonequilibrium charge activation~\cite{Liu2020,Wu2021}. Their main signature is the pinched hysteresis loop in the current-voltage (I-V) response that have been segmented into two main classes~\cite{Pershin2011}: Type I, if a self crossing is present or Type II, with no crossing. However, multiple crossings in the hysteretic memristive loops have been profusely reported in the literature~\cite{Martinez2014,Kubicek2015,Chiolerio2016,Kamble2018,Berruet2022} and they appeared already in the first descriptions of oxide conductance switching~\cite{Hickmott1962,Simmons1967,Argall1968}, way before the memristor concept was coined. 

Scrutinizing their microscopic nature is a challenging task, while understanding the driving mechanisms behind the memory response is a fundamental step for the ability of controlling and tuning them. Thus, it is the purpose of the following description to look beyond the memristive applications and propose the use of the seemingly ubiquitous memory appearance~\cite{Silva2022} as a fundamental characterization tool of transport mechanisms and intrinsic properties of nanostructures and solids in general. In that note, we describe how a full mapping of the transitions of different kinds of responses, ranging from no-crossing to multiple-crossing pinched hysteresis, can become an efficient probing scheme of the concurrence of different transport mechanisms, contrasting natural time scales, and surfacing of symmetry constrains, either extrinsic or intrinsic. The topology modulation of the hysteretic loops, understood here as the character, direction, and number of crossings of the I-V response, will be described within a unified representation, providing clues of its grounding functioning principles on the basis of common elements of transport properties in solids. Thus, the crossing analysis can be extended beyond the Type I and Type II topological dichotomy~\cite{Paiva2022,Silva2022} for the characterization of a wide variety of memeristive responses. A diagrammatic, universal perspective for this characterization in terms of phasors for the time resolved admittance description is also being introduced. For that purpose, the standard phasor analysis of linear ac circuits has been extended to nonlinear responses in a systematic way~\cite{Borys2013,Guo2020}.
\begin{figure}[H]
	\includegraphics[width=8.cm]{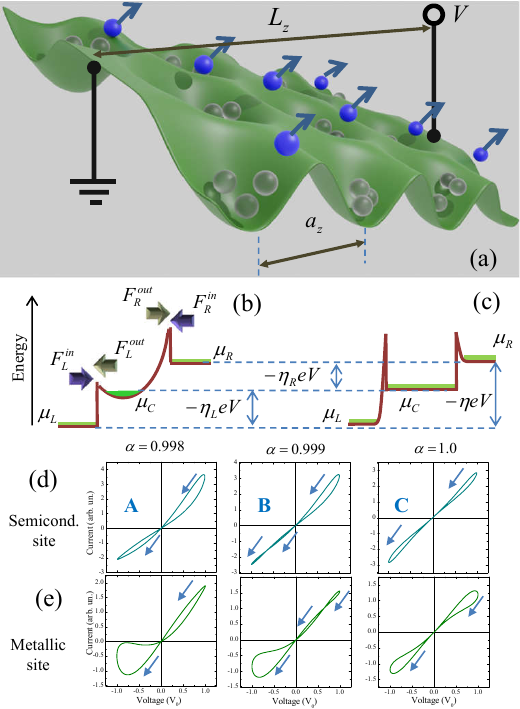}
	\caption{\label{Figure1} (a) Schematic diagram representing the activation of trapped nonequilibrium carriers under external voltage and the local bias determined by the efficiency $\eta=a_z/L_z$. (b) and (c) Diagrams representing trapping sites with metal-semiconductor-metal and semiconductor-metal-semiconductor profiles, respectively. (d) and (e) Evolution of the I-V response for stable  cycles by changing $\alpha$ for semiconducting sites and metallic sites, respectively.}
\end{figure}

\section{Methods}

The mechanism for the conductance switching will be assumed here as triggered by generation and trapping of nonequilibrium carriers (electrons and/or holes), as represented in Figure~\ref{Figure1} (a). For simplicity the model can be projected into one dimension, along the direction of the applied voltage bias, $V$, within a distance, $L_z$, with uniform distribution of localization sites, separated in, $a_z$, onto which symmetry constraints can be further imposed. This allows defining a local voltage drop efficiency as $\eta=a_z/L_z$, directly related with the inverse number of generation sites along the applied voltage direction. 

The following description is a particular example of how to model generation or trapping sites that allows correlating structural parameters and the nonequilibrium carrier dynamics. Thus, for those readers uninterested in these details we can suggest continue reading from the approximation introduced in Eq.~\ref{genrate} that can be applied to a vast universe of generation functions and where the more general discussion resumes.

The carrier traps will be emulated by a  potential profile tuned with the external bias and two configurations as contrasted in Figures~\ref{Figure1} (b) and (c): metal-semiconductor-metal (m-s-m) or semiconductor-metal-semiconductor (s-m-s), henceforth refered to as semiconductive and metallic, respectively. At each generation site, the mechanism of promoting carriers to conducting states is modelled via thermalized Maxwell fluxes of particles through the barriers, $F_{z}=-\frac{n\langle v\rangle}{4}$.~\cite{Kalashnikov1977} Thus, the incoming (outgoing) fluxes are \begin{equation}
   \label{}
F_{\beta}^{\text {in(out) }} =\frac{4 \pi m^{*}\left(k_{B} T\right)^{2}}{(2 \pi \hbar)^{3}} \exp \left(-\frac{E_{\beta}^{B}-\mu_{\beta(C)}}{k_{B} T}\right), 
\end{equation}
with, $\beta=L,R$, labeling the left or right barrier, respectively. Then, the net fluxes at each side are given by $F_{L}=F_{L}^{\text {out }}-F_{L}^{\text {in }}$ and $F_{R}=F_{R}^{\text {out }}-F_{R}^{\text {in }}$ defining the total generation rate within the volume comprised between the contacts as
\begin{equation}
    \centering
    \label{gener}
    g=\frac{S}{\eta}\left(F_{L}+F_{R}\right)
\end{equation}
assuming that $\mu_{c}$ is the \textit{source} of nonequilibrium carriers, where $S$ represents the cross-section area. The effective barriers, according to Figures~\ref{Figure1} (b) and (c) will be considered being $E_B\equiv E^B_L-\mu_L=E^B_R-\mu_R$ for the m-s-m configuration and $E_B\equiv E^B_L-\mu_C=E^B_R-\mu_C$, for the s-m-s one. Note that the main difference between both schemes are the fixed external or internal barriers for either the m-s-m or s-m-s sites, respectively.
Symmetry constraints can be further introduced by allowing a non-uniform local voltage drop, $\mu_{R}-\mu_{L} \equiv-\eta e V$, so that $\mu_{C}-\mu_{L} \equiv-\eta_{L} e V$ and $\mu_{R}-\mu_{C} \equiv-\eta_{R} e V$, where $\eta_{R}=\eta/(1+\alpha)$ and $\eta_{L}=\eta \alpha /(1+\alpha)$, introducing the parameter $\alpha\equiv \eta_{L}/\eta_{R} \in \left[0,\infty \right)$. This parameter quantifies the local symmetry break, with $\alpha=1$ corresponding to the perfectly symmetric case, when $\eta_{R}=\eta_{L}=\eta/2$. Using this representation within Eq.~\ref{gener}, the generation rate for, m-s-m and s-m-s, can be written, respectively, as
\begin{equation}
    \centering
    \label{ga}
    g_{\pm}=\pm \frac{i_{0}}{\eta}\left[\exp \left(\mp \eta_{L} \frac{e V}{k_{B} T}\right)+\exp \left(\pm \eta_{R} \frac{e V}{k_{B} T}\right)-2\right],
\end{equation}
where the upper and lower sign sequences correspond, to m-s-m and s-m-s configurations, respectively, 
with $i_{0}=\frac{4 \pi m^{*} S\left(k_{B} T\right)^{2}}{(2 \pi \hbar)^{3}} e^{-\frac{E_{B}}{k_{B} T}}$. Thus, changing the character of the trapping site is equivalent to changing the sign of $\eta$. Note that Eq.~\ref{ga} is similar to the Butler-Volmer equation used for emulating  electrode reaction kinetics~\cite{Rubi2003}, allowing for a straightforward extension of this model to various electrochemical processes.
Finally, for low local voltage efficiencies, or in general $|\eta| \frac{e V}{k_{B} T}<<1$, the generation rate can be approximated up to the second order terms with respect to the relative voltage amplitude,
\begin{equation}
    \label{genrate}
    g_{\pm} = \sigma_{o} V \pm \sigma_{e} V^{2}
\end{equation}
with
$\sigma_{o} = i_{0}  \frac{1-\alpha}{1+\alpha} \frac{e}{k_{B} T},$
and
$
    \sigma_{e} =  \frac{\eta }{2} i_{0} \frac{1+\alpha^{2}}{(1+\alpha)^{2}} \left(\frac{e}{k_{B} T}\right)^{2}.
$
Note that the \textit{odd-symmetric} generation component with respect to voltage inversion, proportional to $\sigma_{o}$, exactly cancels out for $\alpha=1$. However, the \textit{even-symmetric} component, proportional to $\sigma_{e}$, explicitly depends on $\eta$ and is unavoidable. Since the reciprocal transformation, $\alpha \rightarrow 1/\alpha$, is equivalent to changing the voltage polarity in Eqs. \ref{ga} and \ref{genrate}, we will limit the discussion to $\alpha \in \left[0,1 \right]$.

The time evolution of the number of carriers, $n$, under sweeping voltage, $V=V_0 \cos(\omega t)$, can now be obtained in the relaxation time approximation by solving,
\begin{equation}
\label{master}
dn/dt=g_{\pm}(V)-n/\tau, 
\end{equation}
with $\tau$ representing the nonequilibrium carrier life time. Although this charge dynamics may affect in similar ways a wide range of transport responses, we choose for simplicity to focus in the characterization of a perturbed, Drude-like, linear response. In this case, by assuming a uniform voltage drop between contacts separated by a distance $L_{z}$, the current is $I=\left(G_{0}+\gamma \, n\right) V$, where $G_0$ is the conductance of the unperturbed state~\cite{Silva2022} and $\gamma=e \mu /L_z$, proportional to the carrier mobility $\mu$ ~\cite{Paiva2022}. In the steady regime, (large time limit) which is independent on the initial conditions, the solution is, 
\begin{equation}
\label{it}
I(t)=G_{0} V(t)+I_o(t)+I_e(t),
\end{equation}
with
\begin{equation}
    \centering
    \label{io}
    I_o(t)=\chi V_{0} \frac{\left( 1- \alpha \right)}{\left( 1+\alpha \right)} \frac{ 1}{\left(1+p^2\right)}\left[ 1+\cos \left(2 \omega t \right)+p \, \sin \left(2 \omega t \right)\right],
\end{equation}
$p=\omega \tau$, $\chi=\gamma \frac{i_0}{2} \frac{eV_0}{k_{B} T} \tau$, that condense both intrinsic and extrinsic parameters, while 
\begin{equation}
    \centering
    \label{ie}
    \begin{aligned}
&I_e(t)=\chi V_{0} \nu \frac{\left( 1+ \alpha^2 \right)}{\left( 1+\alpha \right)^2} \frac{1}{2}\left\{ \cos \left(\omega t \right)+ \frac{1}{\left(1+4p^{2}\right)} \cdot \right.\\
&\left. \left[p \, \sin (\omega t)+\frac{\cos (\omega t)}{2}+\frac{\cos (3 \omega t)}{2} +p \, \operatorname{sin}(3 \omega t)\right] \right\},
\end{aligned}
\end{equation}
where $\nu=\eta \frac{eV_0}{k_B T}$ quantifies a relative amplitude efficiency.
Notice that modes up to third order become relevant. Although extending this procedure to higher modes, by considering higher order terms in the expansion of the generation rate in Eq.~\ref{genrate}, is straightforward, the current approximation will already provide a very general picture to describe the conductance tuning with symmetry and external driving parameters. 

\section{Results and Discussion}

This single source of nonequilibrium carriers is a sufficient condition for the emergence of a memristive behavior, as represented in Figures~\ref{Figure1} (d) and (e) by varying the symmetry control parameter $\alpha$ for the two non-equivalent configurations. The contrast with the full numerical result, with no restrictions in Eq.~\ref{ga} with respect to the efficiency, voltage amplitudes, and temperature can be partially found in Ref.~\citenum{Silva2022}. The relative weight of each component described in Eqs.~\ref{ie} and~\ref{io} depends on both intrinsic parameters, such as symmetry constrains and voltage efficiencies; and extrinsic ones, such as voltage amplitudes, temperature, and driving frequencies.

\begin{figure}[H]
	\includegraphics[width=8.6cm]{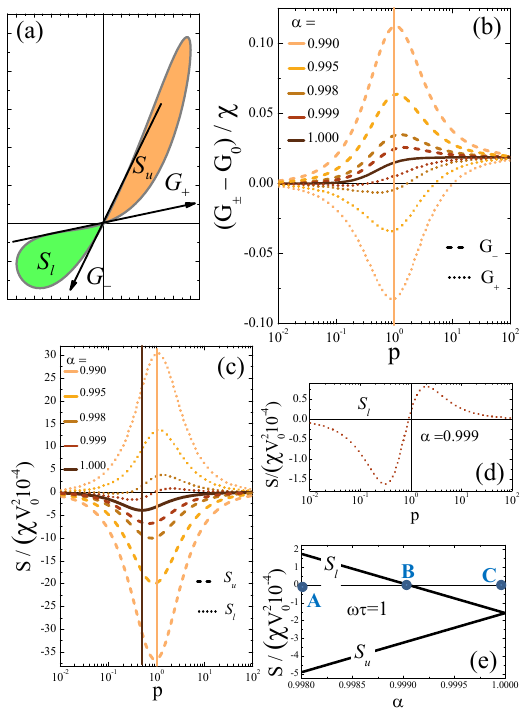}
	\caption{\label{Figure2} (a) Pinched hysteresis and notation for upward and downward conductance at $V=0$ and for upper and lower loop areas. (b) Conductance change with respect to the equilibrium value for upward and downward conductance. (c) Upper and lower loop areas:  The Type II line ($\alpha=1$) has been highlighted with the maximum absolute area at $\omega \tau=1/2$, while the Type I limits for $\alpha \neq 1$ converge towards the maximal area at $\omega \tau=1$. (d) An asymmetric, $\alpha \neq 1$, has been singled out showing a sign reversion of the loop direction by changing frequency. A null area is an indicator of a multiple crossing. (e) Content profiles along the line $\omega \tau=1$ for upper and lower loop areas as functions of $\alpha$ where the labels ABC correspond to the I-V curves in Figure \ref{Figure1} (d).}
\end{figure}

Let us first analyze two limiting cases. For $\alpha=1$, the single contribution of $I_e(t)$ in Eq.~\ref{ie} to the total current results in I-V pinched hysteresis with no crossing, represented in the right panels of Figures~\ref{Figure1} (d) and (e). The contribution of $I_o(t)$ by itself delivers a Type I response, that correspond to the panels at the left of Figures~\ref{Figure1} (d) and (e). The crossing type at zero voltage can be better qualified by obtaining the conductance, $G(t)=I(t)/V(t)$, for $V=0$, in the steady regime, for the down, $G\left(\frac{\pi}{2 \omega} \right)= G_{-}$ and up voltage sweep, $G\left(\frac{3 \pi}{2 \omega} \right)= G_{+}$, with
\begin{equation}
    \centering
    \label{}
    \begin{aligned}
     G_{\pm}  =  
     & G_{0}+\chi \left[ \mp \frac{2 p}{1+p^{2}}\frac{\left( 1- \alpha \right)}{\left( 1+\alpha \right)} + \right.\\
    & \left.  \nu \frac{2 p^{2}}{1+4 p^{2}} \frac{\left( 1+\alpha^2 \right)}{\left( 1+\alpha \right)^2} \right].
    \end{aligned} 
\end{equation}
These values are represented by arrows in Figure~\ref{Figure2} (a). The relative conductance tuning with respect to the equilibrium values $G_0$, $\Delta G_{\pm} \equiv G_{\pm}-G_0$, has been mapped in Figures~\ref{Figure2} (b) and (c) as a function of the relative frequency, $p=\omega\tau$, and the symmetry control, $\alpha$. Note that it is just in the case of $\sigma_o \neq 0$ ($\alpha \neq 1$) that $\Delta G_{+}$ and $\Delta G_{-}$ may differ, leading to the contrasting 'on-off values' of a Type I response. For any combination of parameters, the 'on-off value' difference,
\begin{equation}
    \centering
    \label{}
     G_{-}-G_{+} =  
     \chi \frac{4 p}{1+p^{2}}\frac{\left( 1- \alpha \right)}{\left( 1+\alpha \right)}, 
\end{equation}
is a figure of merit of this model~\cite{Silva2022} that peaks at $p=1$ or $\omega =\tau^{-1}$, and can be used to determine the lifetime if $\sigma_o \neq 0$. In turn, an even symmetric Type II hysteresis is produced for a leading contribution of the term proportional to $\sigma_e$ when $\alpha=1$, when both $G_{-}$ and $G_{+}$ lines coincide in Figures~\ref{Figure2} (b).
\begin{equation}
    \centering
    \label{}
     G_{-}=G_{+}= G_{0}+\chi  \nu \frac{2 p^{2}}{1+4 p^{2}} \frac{\left( 1+\alpha^2 \right)}{\left( 1+\alpha \right)^2}.
\end{equation}
Note, that at intermediary conditions, in the middle panels of Figures~\ref{Figure1} (d) and (e), when both even and odd contributions are balanced, a second crossing of the hysteresis, either on the first or third quadrant of the I-V plane, may emerge. Such a process can be quantified by the content, defined as the area enclosed by the I-V curves, that measures the robustness of the memory response, as represented in Figure~\ref{Figure2} (a). It can be analytically calculated for the steady regime, yielding $S_u=S_{+}$ and $S_l=S_{-}$, for the upper and lower loops, respectively, where
\begin{equation}
    \centering
    \label{su}
    S_{\pm}=-\chi V_0^2 \left[\pm \frac{4}{3} \frac{\left(1-\alpha \right)}{\left( 1+ \alpha \right)} \frac{p}{1+p^{2}}+\frac{\pi}{4} \nu \frac{\left(1+\alpha^2 \right)}{\left( 1+ \alpha \right)^2}\frac{p}{1+4p^{2}}\right].
\end{equation}
The absolute values and sign of the content of the upper and lower loops provide information on the topology of the I-V hysteresis. According to Eqs.~\ref{su}, the maximum content value of both upper and lower loops shifts from $p=1$ ($\omega =1/\tau$), when odd contribution prevails, towards $p=1/2$ ($\omega =1/2\tau$), for leading even term, as displayed in Figure~\ref{Figure2} (c). Notice that this picture changes by switching the nature of the generation sites, $\nu \rightarrow -\nu$ ($\eta \rightarrow -\eta$), with the $\alpha=1$ curve appearing with opposite sign, corresponding to an inverted loop direction. The emergence of an additional crossing beyond $V=0$  eventually turns the upper or lower loop area to zero, while a sign change, in either the first or third quadrant, points to a reversion of the loop direction, according to the generation site profile. This has been represented in Figure~\ref{Figure2} (d) for varying $p=\omega \tau$. The way the topology of the hysteresis changes by varying $\alpha$ can be assessed by correlating the states A, B, C in Figure~\ref{Figure2} (e) with the I-V curves in Figure \ref{Figure1} (d) where the same labels have been used. The local direction of the loop is indicated with arrows.


A geometric representation of the memory response may provide an intuitive meaning to the amplitudes, frequencies, and phase shifts induced by the nonequilibrium charge dynamics and the relative contribution of each generation term. A phasor picture, in particular, allows segmentation of different contributions pointing to the relative weight of each. For that, we shall extend the voltage pulse into the complex plane through the transformation, $V(t) = V_{0} \cos(\omega t) \rightarrow \tilde{V}(t) = V_{0} \, e^{i \omega t}$. A phasor analysis~\cite{Guo2020} can thus be built for a complex current, from which the previous values of Eq.~\ref{it} can be reconstructed. For these goals the following transformations were introduced: $\cos(k \omega t) \rightarrow e^{i k \omega t}$ and $\sin(k \omega t) \rightarrow -i e^{i k \omega t}$, with $k=1,2,$ or $3$. This allows defining a complex admittance, $\tilde{\theta}$, so that $\tilde{I}=\tilde{\theta} \tilde{V}$, with $\tilde{\theta}=G_0+\tilde{\theta}_e+\tilde{\theta}_o$. The contributions, up to the second order, of the even component is
\begin{equation}
\centering
    \label{adme}
\begin{aligned}
    \tilde{\theta}_{e} = \frac{\chi \nu}{2} \frac{\left( 1+\alpha^2 \right)}{\left( 1+ \alpha \right)^2} \left\{\left[1+\dfrac{1}{2\left(1+4p^{2}\right)}\right]-i\dfrac{p}{1+4p^{2}}
     \right. + \\
    \left. \dfrac{1}{2 \sqrt{1+4p^{2}}} \left[ \cos (2 \omega t-\Delta)+i \, \sin(2 \omega t-\Delta) \right] \right\}
\end{aligned}
\end{equation}
with $\Delta=\arctan\left(2 p\right)$, while the odd contribution reads
\begin{equation}
    \label{admo}
    \tilde{\theta}_o=\chi  \frac{\left( 1-\alpha \right)}{\left( 1+ \alpha \right)}\frac{1}{1+p^2} \left[\sqrt{4 + p^{2}} \cos (\omega t-\delta) -i  p  \cos (\omega t)  \right],
\end{equation}
where $\delta=\arctan\left(p/2\right)$.
\begin{figure}[H]
	\includegraphics[width=8.6cm]{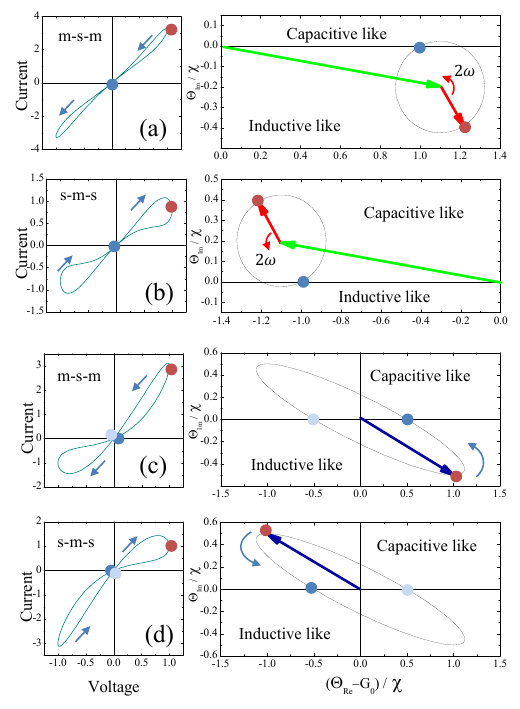}
	\caption{\label{Figure3} Stable I-V cycles on the left and phasor representations for the varying mem-admittance on the right. (a) and (b) Pure even contribution, $\sigma_o=0$ (Type II) for m-s-m and s-m-s configurations, respectively. (c) and (d) Pure odd contribution, $\sigma_e=0$ (Type I) for m-s-m and s-m-s configurations, respectively.}
\end{figure}

The admittance mapping of pure Type II and Type I responses has been displayed in Figure~\ref{Figure3}. By setting $\sigma_o=0$, the even component contribution of the admittance with respect to the unperturbed value, $\tilde{\theta}_e-G_0$, has been highlighted in the right panels of Figures~\ref{Figure3} (a) and (b). Two non-equivalent configurations have been represented that lead to the noncrossing hysteresis of the left panels. The green non-rotating phasors correspond to the complex term on the fist line in Eq.~\ref{adme}, while the $2 \omega$ rotating ones, in red, respond to the second line. Red and blue dots represent the same state in both the I-V curve and the corresponding mem-admittance path. The blue dot, at $V=0$, corresponds to a point of zero susceptance and equal conductance that is attained twice during a full cycle. The generation site configuration (semiconductive or metallic) defines whether the real part of the admittance lies permanently above or below $G_0$. 
Note also that the oscillatory imaginary component of the admittance switches sign within certain fragments of the cycle. Along the positive stretches, the current leads the voltage (in terms of phase-shift), corresponding to a capacitive like susceptance. In turn, negative stretches lead to an inductive-like current that lags the voltage, although no real magnetic induction is present. It is also possible to translate such a behavior in terms of apparent `negative capacitances'~\cite{Bisquert2006,Yadav2019,Gonzales2022,Munoz-Diaz2022} that can be related to the inertia of the charging and discharging processes~\cite{Ershov1998}.
By making $\sigma_e=0$, and leaving just the contribution of Eq.~\ref{admo} to the admittance, the picture corresponds to a Type I response as shown in Figures~\ref{Figure3} (c) and (d). In this case, the response switches between capacitive like lead and inductive like lagging of the current for each semi-period while, as expected, the real-part of the admittance (conductance) attains two contrasting values at $V=0$ during the up and down sweeps, as highlighted with dark and light blue dots, respectively.

The apparent current lagging or leading the voltage are consequences of charging or trapping inertia (retardation) of nonequilibrium carriers. For the pure odd contribution to the generation function, with $\sigma_e=0$, in Figures.~\ref{Figure3} (c) and (d), the trapping or detrapping character is switched by the driving polarity, according to Eq. 4: for one polarity, when $\sigma_o V>0$, the charging inertia provokes a current lagging, interpreted here as an inductive like response, whereas a trapping inertia manifests with the opposite polarity, $\sigma_oV<0$, yielding an apparent current lead, labeled as capacitive like. The case of pure even contribution, when $\sigma_o=0$, displayed in Figures.~\ref{Figure3} (a) and (b) is less trivial since in this case the sign of the generation function does not change along the whole voltage sweep, however its intensity does. Thus, even though the behavior in this case can be either defined as almost capacitive or inductive like (according to the sign of $g_{\pm}$), the trapping or generation efficiency is modulated along each semi-period so that during a fraction of it the apparent current lagging or leading is inverted.

\begin{figure}[H]
	\includegraphics[width=8.6cm]{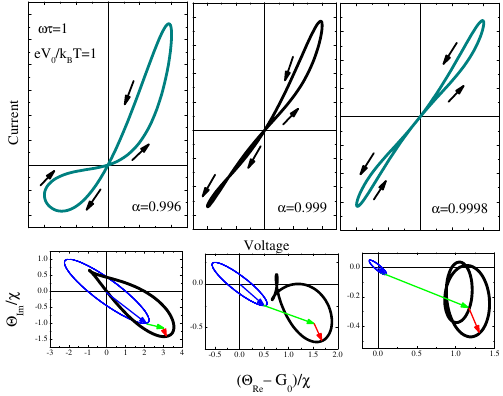}
	\caption{\label{Figure4} Topology tuning of the stable I-V response (top) and the mem-admitance path (bottom) for a single voltage period: fixing $\omega \tau=1$ and varying the symmetry control, $\alpha$.}
\end{figure}

\begin{figure}[H]
	\includegraphics[width=8.6cm]{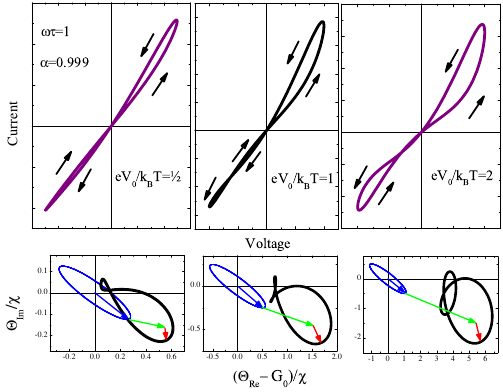}
	\caption{\label{Figure5} Topology tuning of the stable I-V response (top) and the mem-admitance path (bottom) for a single voltage period: fixing $\omega \tau$ and $\alpha$ and varying the voltage amplitude to temperature ratio, $e V_0/ k_B T$.}
\end{figure}

When the odd and even components of a single channel intermix, the admittance path is a linear combination of the pictures described above. In Figure~\ref{Figure4} one may see the topology tuning of the  conductance by changing the symmetry control parameter, $\alpha$. The corresponding mapping of the admittance has been displayed at the bottom of each I-V curve showing how the picture evolves from the odd, Type I, towards an even, Type II response, as $\alpha \rightarrow 1$, passing through a multiple (two) crossing intermediary state (within the third quadrant for $\nu >0$). Note that being a consequence of the intermixing of odd and even components of the current in Eqs.~\ref{io} and \ref{ie}, the multiple crossing picture is not just triggered by intrinsic symmetry constraints. It can also be controlled externally by tuning the driving frequency, as previously illustrated in Figure~\ref{Figure2} (d) or by changing the ratio between the voltage amplitude and temperature, as displayed in Figure~\ref{Figure5}. Thus, the presence of a single channel for nonequilibrium charge generation, combining odd and even contributions, may transit from a Type II response, with zero I-V crossing, to a Type I, with a single crossing at $V=0$, passing through a condition of double crossing (this includes $V=0$). The additional crossing will be located at the upper or lower loop according to the nature and symmetry of the generation sites.

\begin{figure}[H]
   \centering
	\includegraphics[width=8cm]{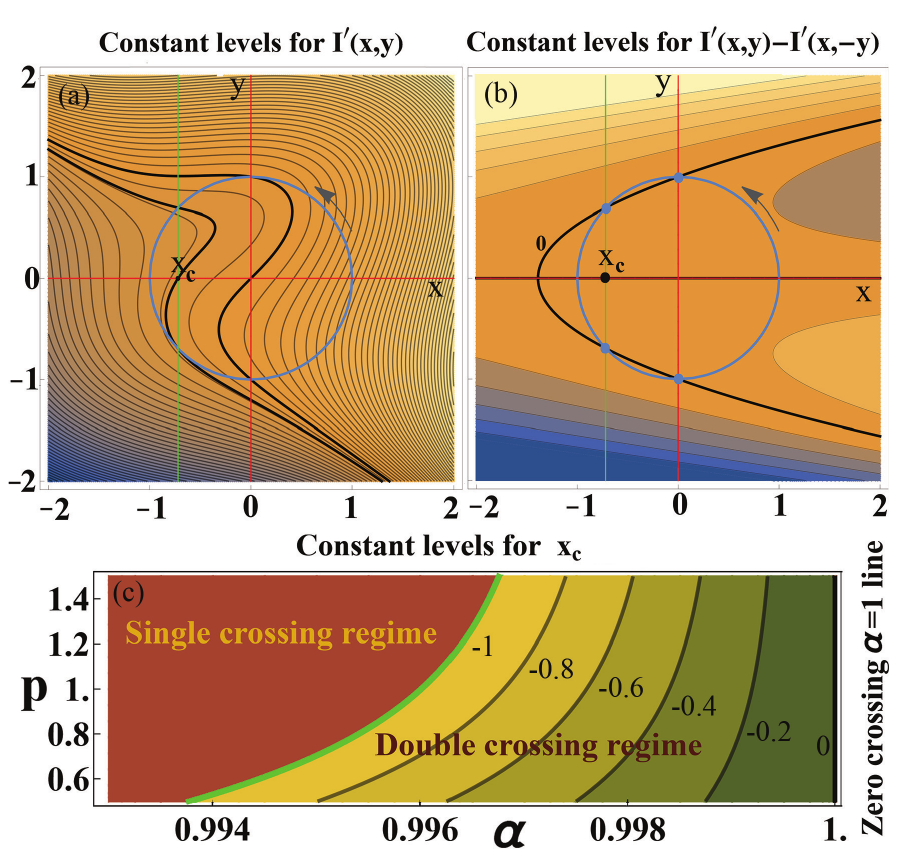}
	\caption{\label{Figure6} Isocurrent cubic curves and the crossing phase diagram. (a) The unit circle corresponds to the voltage path. The two enhanced isocurrents correspond to $x=0$ and $x=x_c\neq0$ crossings, respectively. These are the only isocurrents that cross the unit circle in two conjugate points $(x,\pm y)$. (b) Equivalent representation for the $I'(x,y)-I'(x,-y)$ curves, where hysteresis self-crossing points appear at the intersection of the zero level parabola with the unit circle. (c) Phase diagram for the I-V crossing regimes as a function of the symmetry control parameter $\alpha$ and frequency $p=\omega\tau$.}
\end{figure}

In order to find the number of crossings and their positions in the hysteresis loop we rewrite Eqs.~\ref{io} and \ref{ie} by introducing, $x=V/V_0=\cos(\omega t)$ and  $y=\frac{\tau}{p V_0}\frac{dV}{dt}=-\sin(\omega t)$,  as
\begin{equation}
	I_o=\chi V_0 \frac{\left( 1- \alpha \right)}{\left(1+ \alpha \right)}\frac{2}{\left( 1+p^2 \right)}\left(x^2-p xy \right),
\end{equation}
and
\begin{equation}
	I_e= \chi V_0\nu \frac{\left( 1+ \alpha^2 \right)}{\left( 1+ \alpha \right)^2}\frac{1}{\left( 1+4p^2 \right)}\left(2p^2x+x^3-2py+2py^3 \right).
\end{equation}
Then, the total current can be normalized to $\chi V_0$ and extended to the $(x,y)$ plane yielding
\begin{gather}
	I^{'}(x,y)\equiv \frac{I}{\chi V_0}=g' x+	\frac{1-\alpha}{1+\alpha}\frac{2}{\left(1+p^2\right)}(x^2- p x y)\\\nonumber
	\nu\frac{1+\alpha^2}{(1+\alpha)^2}
	\frac{1}{1+4 p^2}\left(2 p^2 x+x^3-2 p y+2 p y^3\right),
\end{gather}
where $g'=G_0/\chi$. As represented in Figure~\ref{Figure6} (a), the constant level (isocurrent) contours, $I^{'}(x,y)=\text{const}$, are cubic plane curves, which are crossed by a point that moves on the unit circle $x^2+y^2=1$ as a voltage cycle is completed. A sequel representation has been displayed in Figure~\ref{Figure6} (b) where the curves correspond to $I^{'}(x,y)-I^{'}(x,-y)=\text{const}$.
The points on the circle $x^2+y^2=1$ that cross the contour
$I^{'}(x_c,y)-I^{'}(x_c,-y) =0$
correspond to the crossing points of the hysteresis, and beyond the trivial solution, $x_c=0$, a second crossing may appear at 
\begin{gather}
	x_c=-\frac{\left(1-\alpha ^2\right) \left(1+4 p^2\right)}{ \left(\alpha ^2+1\right) \nu  \left(1+p^2\right)}.
\end{gather}
This allows setting a phase diagram for the regimes of zero, $\alpha=1$ (tangential coincidence at $V=0$), one ($|x_c|\geq 1$), or two crossings of the memristive hysteresis ($|x_c|<1$), as represented in Figure~\ref{Figure6} (c). Two is the maximum value of crossings expected with a single generation mechanism. The zero-crossing condition corresponds to the $\alpha=1.0$ line that comprises the Type II responses. The appearance of one or two crossings depends on a combination of symmetry constraints, $\alpha$, relative frequency, $p=\omega \tau$, and effective amplitude to temperature ratio, $\nu=\eta e V_0 / (k_B T)$.

\begin{figure}[H]
   \centering
	\includegraphics[width=8cm]{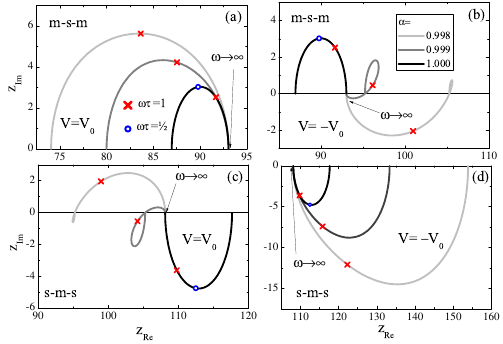}
	\caption{\label{Figure7} Nyquist plots for $\omega \in (0,\infty)$ and varying symmetry constraints, $\alpha$. (a), (b) For a single kind of semiconductive sites and opposite polarities, and analogously for single metallic sources in panels (c), (d). The states corresponding to $\omega \tau=1$ and $1/2$ have been marked by crosses and circles for reference.}
\end{figure}

Let us finalize the discussion by assessing how the nonlinearity of the transport response due to the availability of local generation and/or trapping sites affects the frequency tuning of the conductance and its character. An impedance analysis is best suited for this purpose since it is a state of the art tool for characterizing the performance of transport devices under external oscillatory drives ~\cite{Roman2022}.
However, using impedance spectroscopy nontrivial behaviors, such as the appearance of effective negative capacitance and unexpected complex loops are very common.~\cite{Bou2020,Joshi2020,Hoffmann2021,Srivastava2022} 
More specifically, the emergence of conductive inertia, when the current alternatively lags or leads the voltage, with apparent negative and positive capacitance, has already been experimentally correlated to current-voltage hysteresis.~\cite{Ebadi2019,Gonzales2022}
We should point that these maps, extracted at the maximum voltage amplitude, correspond to a single point in time of the previously defined admittance, which is time dependent. Thus, they must not be mistaken with the spectral impedance analysis of a small signal that in case of nonlinear responses would unavoidably demand a multimode approach.
Here, we illustrate these effects from a microscopic perspective, using the phasor analysis previously introduced. The availability of a single nonequilibrium carriers' channel induces complex admittance responses, which is model dependent and translates into Nyquist maps that become very sensitive to both intrinsic and extrinsic factors. Since the conductance varies with time, we will analyze the response at the voltage cusps, $V=\pm V_0$, spanning the whole frequency domain, $\omega \in (0,\infty)$. The normalized impedance, $Z=\chi/\tilde{\theta}$ ($\chi$ was introduced after Eq. \ref{io}), under stable cycle conditions, becomes, according to Eqs.~\ref{adme} and \ref{admo}, 
\begin{equation}
\begin{aligned}
\frac{1}{Z(\pm V_0)}=g' \pm   \frac{2}{1+p^2}\frac{1-\alpha}{1+\alpha} +\nu \frac{1+2p^2}{1+4p^2}\frac{1+\alpha^2}{\left( 1+\alpha \right)^2} + \\
- i \left[\pm\frac{p}{1+p^2}\frac{1-\alpha}{1+\alpha} +\nu \frac{p}{1+4p^2}\frac{1+\alpha^2}{\left( 1+\alpha \right)^2} \right],
\end{aligned}
\end{equation}
with collapsing susceptance, Im[$\tilde{\theta}(\pm V_0)$]=0, as $\omega=0$ or $\omega \rightarrow \infty$. Note that the resulting Nyquist maps are displayed in Figures~\ref{Figure7} (a)-(d). The outcomes are polarity dependent only if $\alpha \neq 1$, defining a low frequency limit condition, $\omega=0$, $1/Z(\pm V_0)=g' \pm 2 (1-\alpha)/(1+\alpha)+\nu (1+\alpha^2)/(1+\alpha)^2$.
There is in turn an absolute limit of the impedance for any configuration at high frequencies, since $\lim_{p\rightarrow \infty} 1/Z(\pm V_0)=g'+\nu/2 (1+\alpha^2)/(1+\alpha)^2$. The states corresponding to the optimal drives unveiled by Eq.~\ref{su} and Figure~\ref{Figure2} (c), $p=1$ and $p=1/2$, have been pinpointed on the curves for reference.

In the pure even case, with $\alpha=1$, the maps coincide in both Figures~\ref{Figure7} (a) and (b); and (c) and (d). In Figures~\ref{Figure7} (a) and (b), the system responds as a permanent carrier source ($G>G_0$), for the semiconducting site configuration ($\nu>0$), with a resistance that increases as $\omega \rightarrow \infty$, and an inductive current lagging (negative capacitance) for intermediary frequencies. Meanwhile, the metallic site ($\nu<0$), illustrated in Figures~\ref{Figure7} (c) and (d), would act as a permanent carrier drain ($G<G_0$), with capacitive like character, and a resistance decrease as $\omega \rightarrow \infty$. 
In turn, the introduction of odd contributions, for $\alpha \neq 1$, may either boost this behavior depending on the voltage polarity and site character, as demonstrated in Figures~\ref{Figure7} (a) and (d), or counterbalance it, as displayed in Figures~\ref{Figure7} (b) and (c). In this latter case, the loops can spiral, switching from inductive to capacitive behaviors by changing the frequency, according to internal symmetry constrains and the values of $\nu$. The odd component forces the system to switch the character between drain and source of nonequilibrium carriers, as the real-part of the admittance oscillates around $G_0$ by switching $V=V_0$ to $V=-V_0$, as illustrated in Figures~\ref{Figure3} (c) and (d).

\section{Conclusions}

In summary, the availability of generation or trapping sites for nonequilibrium charge carriers is able to produce a memory response under the adequate set of driving inputs. A single channel and external factors such as pulse frequencies, amplitudes, and temperature, when combined with internal symmetry constraints allow for the tuning of the conductance to produce regimes with zero, one, or two crossings of the memristive hysteresis. The appearance of more than two crossings in the hysteresis would point unequivocally to the concurrence of more than one channel. As a result, the conjecture raised in Ref.~\citenum{Kubicek2015} of two competing switching mechanisms triggering their triple crossing response has been theoretically confirmed. This analysis suggests that, when the conductance state is modulated by nonequilibrium carriers, it may combine capacitive and inductive like elements that shift the current with respect to the voltage according to the nature of the generation site and produce complex impedance maps. It has also been demonstrated that the complexity of these plots is not necessarily an outcome of combining complex microscopic elements. Besides, in the presence of more than one channel, this impedance mapping can be a tool for elucidating the generation sites' character, contrasting their natural time scales, and assessing their relative impact in the transport response. These findings can be extended to a wide range of electrochemical conduction switching effects where the emergence of memory features is pervasive. Although the generation function is written explicitly for non-equilibrium carriers from trapping centers, the activation process underlying this function can be transferred to other processes, such as to e.g. those driven by the Butler-Volmer equation that emulates electrode reaction kinetics.

\section{Acknowledgments}

This study was financed in part by the Conselho Nacional de Desenvolvimento
Científico e Tecnológico - Brazil (CNPq) Projs. 301033/2019-6 and 131951/2021-1, and the Fulbright Program of the United States Department of State’s
Bureau of Educational and Cultural Affairs.

\newpage









\section[here]{Data Availability Statement}

The data that supports the findings of this study are available within the article.


\bibliography{resubJAP.bbl}

\end{document}